\documentclass[10pt]{article}
\usepackage[english]{babel}
\usepackage[T1]{fontenc}
\usepackage{latexsym}
\usepackage{amsfonts}
\usepackage{epsfig}
\usepackage{dsfont}
\usepackage{cprform}

\begin{document}
\def\op#1{\mathcal{#1}}
\def\bfnull{\relax{\rm O \kern-.635em 0}}
\def\dop{{\rm d}\hskip -1pt}
\def\a{\alpha}
\def\b{\beta}
\def\g{\gamma}
\def\d{\delta}
\def\e{\epsilon}
\def\ve{\varepsilon}
\def\t{\theta}
\def\l{\lambda}
\def\m{\mu}
\def\n{\nu}
\def\pg{\pi}
\def\r{\rho}
\def\s{\sigma}
\def\t{\tau}
\def\z{\zeta}
\def\c{\chi}
\def\p{\psi}
\def\o{\omega}
\def\G{\Gamma}
\def\D{\Delta}
\def\T{\Theta}
\def\L{\Lambda}
\def\Pg{\Pi}
\def\S{\Sigma}
\def\O{\Omega}
\def\pb{\bar{\psi}}
\def\cb{\bar{\chi}}
\def\lb{\bar{\lambda}}
\def\Pii{\mathcal{P}}
\def\Q{\mathcal{Q}}
\def\K{\mathcal{K}}
\def\A{\mathcal{A}}
\def\N{\mathcal{N}}
\def\F{\mathcal{F}}
\def\Gi{\mathcal{G}}
\def\Ci{\mathcal{C}}
\def\oL{\overline{L}}
\def\oM{\overline{M}}
\def\wk{\widetilde{K}}
\def\hb{\overline{h}}
\def\eq#1{(\ref{#1})}
\newcommand{\be}{\begin{equation}}
\newcommand{\ee}{\end{equation}}
\newcommand{\ba}{\begin{eqnarray}}
\newcommand{\ea}{\end{eqnarray}}
\newcommand{\ban}{\begin{eqnarray*}}
\newcommand{\ean}{\end{eqnarray*}}
\newcommand{\nn}{\nonumber}
\newcommand{\nin}{\noindent}
\newcommand{\fgl}{\mathfrak{gl}}
\newcommand{\fu}{\mathfrak{u}}
\newcommand{\fsl}{\mathfrak{sl}}
\newcommand{\fsp}{\mathfrak{sp}}
\newcommand{\fusp}{\mathfrak{usp}}
\newcommand{\fsu}{\mathfrak{su}}
\newcommand{\fp}{\mathfrak{p}}
\newcommand{\fso}{\mathfrak{so}}
\newcommand{\fg}{\mathfrak{g}}
\newcommand{\fr}{\mathfrak{r}}
\newcommand{\fe}{\mathfrak{e}}
\newcommand{\rE}{\mathrm{E}}
\newcommand{\rSp}{\mathrm{Sp}}
\newcommand{\rSO}{\mathrm{SO}}
\newcommand{\rSL}{\mathrm{SL}}
\newcommand{\rSU}{\mathrm{SU}}
\newcommand{\rUSp}{\mathrm{USp}}
\newcommand{\rU}{\mathrm{U}}
\newcommand{\rF}{\mathrm{F}}
\newcommand{\R}{\mathbb{R}}
\newcommand{\C}{\mathbb{C}}
\newcommand{\Z}{\mathbb{Z}}
\newcommand{\Hb}{\mathbb{H}}
\def\oL{\overline{L}}
\def\mW{\mathcal{W}}


\begin{titlepage}
\begin{center}

\rightline{\small IFT-UAM/CSIC-07-29}
\vskip 1cm

{\Large \bf  Domain Wall/Cosmology correspondence \\ in $(AdS/dS)_6\times S^4$ geometries}
\vskip 1.2cm

{\bf 
Silvia Vaul\`a }
\vskip 0.2cm
{\it Instituto de F\'{\i}sica Te\'orica UAM/CSIC\\
Facultad de Ciencias C-XVI,  C.U.~Cantoblanco,  E-28049-Madrid, Spain}

\vskip 0.4cm

 {\tt silvia.vaula@uam.es} 
\vskip 0.4cm

\end{center}

\vskip 1cm

\begin{center} {\bf ABSTRACT }\end{center}
\vskip 0.4cm

We investigate the ten dimensional origin of six dimensional $F_4$ variant supergravity with supersymmetric  de Sitter background. We address first the issue of spontaneous compactification, showing that it consists of a warped compactification on a four sphere of a variant massive type IIA supergravity. Moreover we illustrate how the known D4-D8 brane solution, whose near horizon geometry yields $AdS_6\otimes S^4$, is accordingly modified to a system including Euclidean branes. Finally, we discuss the relation between this latter solution and  the D4-D8 brane system, showing how it represents a generalisation of the DW/Cosmology correspondence. 

\noindent

\vfill

\end{titlepage}

\section{Introduction}
Among the supergravity theories with supersymmetric $AdS$ vacua, $D=6$ $N=2$, supergravity based on the exceptional supergroup ${\rm F}_4$~\cite{Romans:1985tw}  is somehow peculiar. Indeed,  ${\rm F}_4$ appears to be the  only supergroup admitting two real sections\footnote{See for instance~\cite{VanProeyen:1999ni,Ferrara:2001dn}.}  whose bosonic generators span respectively the algebra ${\rm SO}(2,5)\otimes{\rm SU}(2) $ and ${\rm SO}(1,6)\otimes{\rm SU}(2)$. This reflects into the existence of two version of ${\rm F}_4$ supergravity: the standard one\footnote{We specify for every theory the space--time signature $(t,s)$ in which is formulated.}, ${\rm F}_4(1,5)$, with supersymmetric $AdS_6$ background~\cite{Romans:1985tw,D'Auria:2000ad,Andrianopoli:2001rs}, and a variant version, ${\rm F}^*_4(1,5)$, with supersymmetric $dS_6$ background~\cite{D'Auria:2002fh}.  ${\rm F}^*_4(1,5)$ is a "variant" theory in the sense discussed by Hull~\cite{Hull:1998vg}.  Variant type II supergravities were introduced~\cite{Hull:1998vg}  considering T--duality transformations involving timelike circles; consequently  lower dimensional variant supergravities naturally arise e.g. from compactifications on non--Euclidean tori. 
Hence, variant supergravities can occur in non--Lorentzian signatures.  Quite generally they also have ghosts, and in lower dimensions they may have non--compact $R$--symmetry groups. 

${\rm F}^*_4(1,5)$ supergravity has Lorentzian signature, nevertheless is a variant theory since its vector fields are ghosts. Remarkably the $R$--symmetry group  is compact, since it is ${\rm SU}(2)$ for both real sections. This fact turns out to be quite  relevant in the understanding of its ten dimensional origin.

It is in fact well known~\cite{Cvetic:1999un} that  ${\rm F}_4(1,5)$ supergravity can be obtained from a consistent Kaluza--Klein compactification of massive  ${\rm IIA_m}(1,9)$~\cite{Romans:1985tz} on a four--sphere. More precisely  not on the whole sphere, rather on an hemisphere $\tilde{S}^4$ viewed  as a foliation of three--spheres $S^3$, whose rigid deformations parametrise ${\rm SU}(2)$.  This observation strongly suggests that ${\rm F}^*_4(1,5)$ must come from a similar compactification where at least the foliating $S^3$ is a genuine compact three-sphere.

In Section 1 we will see that this is actually the case.  Modifying the ansatz in~\cite{Cvetic:1999un} we show that ${\rm F}^*_4(1,5)$ can be obtained from a compactification of\footnote{${\rm IIA^*_m}(5,5)$ is the massive version of ${\rm IIA^*}(5,5)$ introduced in~\cite{Hull:1998ym}.}  ${\rm IIA^*_m}(5,5)$ on a timelike four--sphere $\tilde{S}^4$. 
The signature $(5,5)$ is rather peculiar: it is in fact the only signature, together with $(1,9)$, for which  we can impose (pseudo--)Majorana and  Weyl conditions at the same time.  Moreover, apart from the space--time signature, the action of ${\rm IIA_m}(5,5)$ coincides with the action of ${\rm IIA_m}(1,9)$. The same happens for  ${\rm IIA_m^*}(5,5)$ and  ${\rm IIA_m^*}(1,9)$, since the action of both theories exhibits  reversed sign for the kinetic terms of  the RR fields~\cite{Hull:1998ym} and the scalar potential.

Matter coupled ${\rm F}_4(1,5)$  supergravity~\cite{D'Auria:2000ad,Andrianopoli:2001rs}  admits as well an $AdS_6$ supersymmetric background, which has an holographic description in terms of a five dimensional  superconformal field theory~\cite{Ferrara:1998gv}. This result can be interpreted as the correspondence between the near horizon geometry  and world--volume theory  of a system of ${\rm D4-D8}$ branes~\cite{Ferrara:1998gv,Brandhuber:1999np}. \\
It is therefore natural to ask oneself whether a similar holographic description can be found as well for the $dS_6$ vacuum of  matter coupled ${\rm F}^*_4(1,5)$ supergravity~\cite{D'Auria:2002fh}. In particular, given that ${\rm F}^*_4(1,5)$ supergravity can be obtained by compactification of ${\rm IIA^*_m}(5,5)$, we expect the relevant brane system to be a solution of the latter~\cite{Hull:1998vg}.

In Section 2,  we propose a system of D(p,q)--branes  of  ${\rm IIA^*_m}(5,5)$ whose near horizon geometry is actually $dS_6\otimes_w \tilde{S}^4$, with timelike $\tilde{S}^4$, in the same way $AdS_6\otimes_w \tilde{S}^4$ with spacelike $\tilde{S}^4$ arises as near horizon geometry of the ${\rm D4-D8}$ system  of ${\rm IIA_m}(1,9)$~\cite{Brandhuber:1999np}. Moreover the localised  D4--brane of ${\rm IIA_m}(1,9)$  corresponds to an Euclidean E5--brane of ${\rm IIA_m^*}(5,5)$, suggesting that if an holographic description is actually possible, this should correspond to the fixed point of an Euclidean Super Yang--Mills theory~\cite{Hull:1998vg}; hence  the natural candidate would be an Euclidean,  eventually variant,  version of~\cite{Ferrara:1998gv}.

In Section 3 the system of D(p,q)--branes  of  ${\rm IIA^*_m}(5,5)$ is discussed in the context of the Domain--Wall/Cosmology correspondence~\cite{Skenderis:2006fb}. 
In fact, in both ${\rm F}^*_4(1,5)$ and  ${\rm IIA_m^*}(5,5)$ spinors enjoy pseudo--reality conditions. This in turn means that  the aforementioned brane system admits a set of  of pseudo--real Killing spinors.  Similarly the pseudo--Killing spinors of cosmological solutions in the DW/Cosmology correspondence enjoy pseudo--reality conditions.  There is in fact an example~\cite{Bergshoeff:2007cg} of correspondence between a DW solution of ${\rm IIA_m}(1,9)$ and a cosmological solution of ${\rm IIA^*_m}(1,9)$ with pseudo--real Killing spinors, suggesting that pseudo--supersymmetry can be quite generally realised as  supersymmetry of a variant theory. Similarly, the construction of ${\rm F}^*_4(1,5)$~\cite{D'Auria:2002fh} is based on the choice of pseudo--reality condition on the spinors,  contrary to ${\rm F}_4(1,5)$, where reality is imposed~\cite{D'Auria:2002fh}. This kind of relation between supersymmetric AdS and dS vacua  was also recently discussed in the context of DW/Cosmology correspondence in~\cite{Skenderis:2007sm}.  \\
Here we generalise the correspondence between pseudo--supersymmetric solutions of ordinary supergravity theories  and supersymmetric solutions of variant supergravity  theories, providing as a specific example the D(p,q)--branes system  of  ${\rm IIA^*_m}(5,5)$ under discussion.


\section{Spontaneous compactification}

In this section we show, in the same spirit of~\cite{Liu:2003qa}, how  to modify the compactification ansatz of~\cite{Cvetic:1999un} in order to obtain a spontaneous compactification of ${\rm IIA_m^*}(5,5)$ to ${\rm F}^*_4(1,5)$. 

In~\cite{Vaula:2002cn} it was shown how  to map a standard supergravity into a variant supergravity analysing  the properties of the spinors, which depend on the signature of the space--time and if reality or pseudo--reality\footnote{Reality and pseudo--reality  correspond to "convention I" and "convention II" respectively in~\cite{Vaula:2002cn}.}conditions are imposed. The map between  ${\rm F}_4(1,5)$ and ${\rm F}^*_4(1,5)$ was derived in~\cite{D'Auria:2002fh}, while the map between ${\rm IIA}(5,5)$ and ${\rm IIA^*}(5,5)$ was derived in~\cite{Vaula:2002cn}; the generalisation to the massive case can be easily done with the same techniques of~\cite{Vaula:2002cn}.

Let us stress from the very beginning, that for convenience we are going to use the same convention for the metric as in~\cite{Cvetic:1999un,Brandhuber:1999np}, that is 
\be \eta_{(t,s)}=(\underbrace{-,-,-}_{t\,times},\dots\underbrace{+,+,+}_{s\,times})\label{signature}\ee
while in~\cite{D'Auria:2002fh,Vaula:2002cn} we used the opposite one.

Let us start by presenting the bosonic Lagrangians of  ${\rm IIA_m}(1,9)$ in the conventions of~\cite{Cvetic:1999un}
\ba
\mathcal{L}_{1,9}\!\!\!\!&=&\!\!\!\! \hat{R}-\frac12\hat{*}d\hat{\phi}\wedge d\hat{\phi}-\frac12e^{\frac32\hat{\phi}}\hat{*}d\hat{F}_{(2)}\wedge d\hat{F}_{(2)}-\frac12e^{-\hat{\phi}}\hat{*}d\hat{H}_{(3)}\wedge d\hat{H}_{(3)}-\frac12e^{\frac12\hat{\phi}}\hat{*}d\hat{F}_{(4)}\wedge d\hat{F}_{(4)}\nn\\
&&-\frac12 d\hat{A}_{(3)}\wedge d\hat{A}_{(3)}\wedge \hat{B}_{(2)}-\frac16m d\hat{A}_{(3)}\wedge (\hat{B}_{(2)})^2-\frac1{40}m^2(\hat{B}_{(2)})^5-\frac12m^2 e^{\frac52\hat{\phi}} \hat{*}\bf1\label{L10}
\ea
where the field strengths are defined in terms of the potentials according to
\ba
&&\hat{F}_{(2)}=d\hat{A}_{(1)}+m\,\hat{B}_{(2)},\quad\quad\hat{H}_{(3)}=d\hat{B}_{(2)}\nn\\
&&\hat{F}_{(4)}=d\hat{A}_{(3)}+\hat{A}_{(1)}\wedge d\hat{B}_{(2)}+\frac12m\,\hat{B}_{(2)} \wedge\hat{B}_{(2)}\label{fs10}
\ea
The map for going to ${\rm IIA_m^*}(5,5)$ is given by 
\be \hat{A}_{(1)}\rightarrow i \hat{A}_{(1)};\quad \hat{B}_{(2)}\rightarrow -\hat{B}_{(2)};\quad\hat{A}_{(3)}\rightarrow -i \hat{A}_{(3)};\quad m\rightarrow -im\label{map10}\ee
together with a signature redefinition $\eta_{(1,9)}\rightarrow\eta_{(5,5)}$.\\
Note that there is some freedom in the choice of the signs in \eq{map10}, the only relevant thing is the reality of the coefficient in the redefinition.  We made this choice in oder to have homogeneous scaling of \eq{fs10} and for later consistency with the six dimensional map.

The corresponding Lagrangian is easily obtained 
\ba
\mathcal{L}^*_{5,5}\!\!\!\!&=&\!\!\!\! \hat{R}-\frac12\hat{*}d\hat{\phi}\wedge d\hat{\phi}+\frac12e^{\frac32\hat{\phi}}\hat{*}d\hat{F}_{(2)}\wedge d\hat{F}_{(2)}-\frac12e^{-\hat{\phi}}\hat{*}d\hat{H}_{(3)}\wedge d\hat{H}_{(3)}+\frac12e^{\frac12\hat{\phi}}\hat{*}d\hat{F}_{(4)}\wedge d\hat{F}_{(4)}\nn\\
&&-\frac12 d\hat{A}_{(3)}\wedge d\hat{A}_{(3)}\wedge \hat{B}_{(2)}+\frac16m d\hat{A}_{(3)}\wedge (\hat{B}_{(2)})^2-\frac1{40}m^2(\hat{B}_{(2)})^5+\frac12m^2 e^{\frac52\hat{\phi}} \hat{*}\bf1\label{L10*}
\ea
with the same definition \eq{fs10} for the field strengths.

In the notations of~\cite{Cvetic:1999un}, the bosonic Lagrangian of pure ${\rm F}_4(1,5)$ supergravity~\cite{Romans:1985tw}  is 
\ba
\mathcal{L}_{1,5}\!\!\!\!&=&\!\!\!\! {R}-\frac12{*}d{\phi}\wedge d{\phi}-\frac12e^{-\frac1{\sqrt2}{\phi}}({*}d{F}_{(2)}\wedge d{F}_{(2)}+*d{F}^i_{(2)}\wedge d{F}^i_{(2)})-\frac12e^{-\sqrt2{\phi}}{*}d{H}_{(3)}\wedge d{H}_{(3)}\nn\\
&&-\frac12 {B}_{(2)}\wedge (\frac12 d{A}_{(1)}\wedge d{A}_{(1)}+\frac13g {B}_{(2)}\wedge d{A}_{(1)}+\frac2{27}g^2{B}_{(2)}\wedge B_{(2)}+\frac12 {F}^i_{(2)}\wedge {F}^i_{(2)})\nn\\
&&-g^2 (\frac29e^{\frac3{\sqrt2}\phi}-\frac83e^{-\frac1{\sqrt2}\phi}-2e^{-\frac1{\sqrt2}\phi}){*}\bf1\label{L6}
\ea
where the field strengths are defined in terms of the potentials according to
\ba
&&{F}_{(2)}=d{A}_{(1)}+\frac23 g\,{B}_{(2)},\quad\quad{H}_{(3)}=d{B}_{(2)}\nn\\
&&{F}^i_{(2)}=d{A}^i_{(1)}+\frac12 g\,\e_{ijk}{A}^j_{(1)}\wedge{A}^k_{(1)};\quad i=1,\,2,\,3\label{fs6}
\ea
The map for going to ${\rm F}_4^*(1,5)$ is given by
\be (A_{(1)},\, A^i_{(1)})\rightarrow i (A_{(1)},\, A^i_{(1)});\quad B_{(2)}\rightarrow -B_{(2)};\quad g\rightarrow -ig\label{map6}\ee 
and the corresponding Lagrangian 
\ba
\mathcal{L}^*_{1,5}\!\!\!\!&=&\!\!\!\! {R}-\frac12{*}d{\phi}\wedge d{\phi}+\frac12e^{-\frac1{\sqrt2}{\phi}}({*}d{F}_{(2)}\wedge d{F}_{(2)}+*d{F}^i_{(2)}\wedge d{F}^i_{(2)})-\frac12e^{-\sqrt2{\phi}}{*}d{H}_{(3)}\wedge d{H}_{(3)}\nn\\
&&-\frac12 {B}_{(2)}\wedge (\frac12 d{A}_{(1)}\wedge d{A}_{(1)}+\frac13g {B}_{(2)}\wedge d{A}_{(1)}+\frac2{27}g^2{B}_{(2)}\wedge B_{(2)}+\frac12 {F}^i_{(2)}\wedge {F}^i_{(2)})\nn\\
&&+g^2 (\frac29e^{\frac3{\sqrt2}\phi}-\frac83e^{-\frac1{\sqrt2}\phi}-2e^{-\frac1{\sqrt2}\phi}){*}\bf1\label{L6*}
\ea
with the same definitions \eq{fs6} for the field strengths. 

The compactification ansatz of~\cite{Cvetic:1999un} substituted into the ten dimensional equations of motion of \eq{L10} implies the six dimensional equations of motion of \eq{L6}.\\ 
If we apply the map \eq{map6} to the compactification ansatz of~\cite{Cvetic:1999un} we easily realise that the equations of motion of \eq{L6*} imply the equation of motion of \eq{L10*}. 

In order to understand it, let us first consider the equations of motion of the form fields and the dilaton: applying the map \eq{map6} to the ansatz  in~\cite{Cvetic:1999un} it is immediate to see that the map \eq{map10} on the ten dimensional fields is enforced, hence giving  the equations of motion of \eq{L10*}. Some more details are given in Appendix A.

Consider now the Einstein equation and the metric ansatz of~\cite{Cvetic:1999un}
\be ds^2_{10}=h(\z,\phi))[ds^2_6+2\,g^{-2}(f_1(\z,\phi) d\z^2+f_2(\z,\phi) cos^2\z\, h^ih^i)]\label{m10}\ee
where $h^i=\s^i-gA^i_{(1)}$, with $\s^i$ left invariant 1--forms on the 3--sphere $S^3$ and $ds^2_6$ an Einstein metric. For $A_{(1)}^i=\phi=0$ we have that $f_1=1,\,=f_2=\frac14$ . The ten dimensional geometry becomes
\be ds^2_{10}=h(\z,0))[ds^2_6+2g^{-2}(d\z^2+\frac14cos^2\z\, \s^i\s^i)] \label{m6+4}\ee 
where $ds^2_4=d\z^2+cos^2\z\, \s^i\s^i$ is the metric of the four--sphere $\tilde{S}^4$, with Ricci tensor $R_{\a\b}=\frac32 g^2g_{\a\b}$ and $ds^2_6$ is an $AdS_6$ metric with Ricci tensor $R_{ab}=-\frac{10}9g^2g_{ab}$.

Applying the map \eq{map6} to the metric \eq{m10}, the functions $h,\,f_1,\,f_2$ remain untouched, while $g^{-2}\rightarrow -g^{-2}$. This turns  \eq{m10} into
\be ds^{*\,2}_{10}=h(\z,\phi))[ds^2_6-2\,g^{-2}(f_1(\z,\phi) d\z^2+f_2(\z,\phi) cos^2\z\, h^ih^i)]\label{m10*}\ee
The change of sign in the metric ansatz means that four out of the nine spacelike directions have become timelike.  This implies the change of signature $\eta_{(1,9)}\rightarrow\eta_{(5,5)}$ in the ten dimensional theory  and ensures that the Einstein equation is satisfied since the stress--energy tensor of \eq{L10*} is clearly obtained applying the map \eq{map10} to the stress--energy tensor of \eq{L10}.

For $A_{(1)}^i=\phi=0$ the metric \eq{m10*}  becomes 
\be ds^{*\,2}_{10}= h(\z,0))[ds^2_6-2g^{-2}(d\z^2+\frac14cos^2\z\, \s^i\s^i)] \label{m6+4*}\ee
In this case $ds^2_6$ is a $dS_6$ metric with Ricci tensors $R_{\a\b}=-\frac32 g^2g_{\a\b}$, while $ds^2_4$ is still the metric of a four--sphere $\tilde{S}^4$ with Ricci tensor   $R_{ab}=\frac{10}9g^2g_{ab}$. The change of sign of the Ricci tensor of $\tilde{S}^4$  is due to the fact that now the ten dimensional metric has signature $(5,5)$ which splits $(5,5)\rightarrow (1,5)+(4,0)$;  the Lorentzian part $(1,5)$ pertains to $dS_6$, while $\tilde{S}^4$ is completely timelike, hence the Ricci tensor has the opposite sign.\\
As discussed before, the fact that the internal manifold has to be a foliation of three--spheres, is related to the six dimensional $R$--symmetry group, which is ${\rm SU}(2)$ for both ${\rm F}_4(1,5)$ and ${\rm F}_4^*(1,5)$.


\section{Brane solutions and near--horizon geometry}
In this section we will discuss how the D4--D8 system of type ${\rm IIA_m}(1,9)$~\cite{Brandhuber:1999np} which gives as  near horizon geometry a warped product $AdS_6\otimes \tilde{S}^4$ is mapped into a system of branes in ${\rm IIA_m^*}(5,5)$ giving as near horizon geometry a warped product $dS_6\otimes \tilde{S}^4$ with timelike four--sphere.

Let us start with the metric of the D4-D8 brane system~\cite{Brandhuber:1999np} 
\be
d{s}_{10}^2=(H_4 H_8)^{-\frac12}(-dt^2+d\vec{\o}^2)+H_4^{\frac12} H_8^{-\frac12} d\vec{x}^2+(H_4 H_8)^{\frac12} dz^2\label{D8D4}
\ee
where we have defined
\be dt^2=dt_\bullet dt_\bullet,\quad\quad d\vec{\o}^2=\sum_{i=1}^{4}d\o_id\o_i, \quad\quad\vec{x}^2=\sum_{i=1}^{4}dx_idx_i,\quad\quad dz^2=dz_\bullet dz_\bullet\,.\ee
where the indexes (including the dots $\bullet$ which we  introduced for sake of clarity) are raised and lowered with the ten dimensional metric $\eta_{(1,9) }$ \eq{signature}, i.e. $dt_\bullet=-dt^\bullet$, $dx_i=dx^i$, etc. 
The harmonic function $H_8$ depends on the sole coordinate  $z$ which is transverse to the D8--brane, while $H_4$ depends on $(z,\,\vec{x})$ which are transverse to the D4--brane. For  localised D4--branes, they have to satisfy~\cite{Youm:1999zs,Brandhuber:1999np}
\be \partial_z\partial_z H_8(z)=0,\quad\quad\partial_z\partial_z H_4(z,\vec{x})+H_8(x)\sum_{i=1}^{4}\partial_{x_i}\partial_{x_i}H_4(z,\vec{x})=0 \label{interD48}\ee
Performing a change of coordinates $\tilde{z}=\frac23 z^{\frac32}$, the metric can be put into the form 
\be
d{s}_{10}^2=(\tilde{H}_8)^{-\frac12}\left[\tilde{H}_4^{-\frac12} (-dt^2+d\vec{\o}^2)+\tilde{H}_4^{\frac12} (d\vec{x}^2+d\tilde{z}^2)\right]
\ee
where at  the near horizon
\be \tilde{H}_8(\tilde z)=Q_8\left(\frac94\tilde{z}^2\right)^{\frac13};\quad\quad \tilde{H}_4(\tilde{z},\vec{x})=\frac{Q_4}{[x^2+\tilde{z}^2]^{\frac53}}\label{HD48}\ee
Again, we define $x^2=\sum_{i=1}^{4}x_ix_i$, and $\tilde{z}^2=z_\bullet z_\bullet$.

Performing first the change of coordinates $x=r\,sin\a$, $\tilde{z}=r\,cos\a$, $r\geq0$, $0\leq\a\leq\frac\pi2$ and afterwards  defining $r^2=U^3$, together with a  rescaling $\o_i\rightarrow \frac23Q_4^{\frac14}\o_i$ we can put \eq{D8D4} into a convenient near horizon form~\cite{Brandhuber:1999np} 
\be
d\hat{s}_{10}^2=\frac94Q_4^\frac12\left(\frac32Q^{\frac12}_8sin\a\right)^{-\frac13}\left[U^2dy^2_{||}+\frac{dU^2}{U^2}+d\O^2_4 \right]\label{D8D4N}
\ee
where $dy^2_{||}=-dt^2+d\vec{\o}^2$ and $d\O^2_4=d\a^2+cos^2\a\,d\O^2_3$, which coincides with \eq{m6+4}.

We now want to map it into a solution of ${\rm IIA}_m^*(5,5)$. In order to do that, we first transform it into a D--branes solution solution of ${\rm IIA}_m(5,5)$. This is quite immediate, since the action of  ${\rm IIA}_m(5,5)$ coincides with the standard action ${\rm IIA}_m(1,9)$, a part from the space--time signature. Since the timelike directions belong to the world--volume of D--branes, we have just to change the signature, that is  $d\vec{\o}^2\rightarrow -d\vec{\o}^2$, which doesn't affect the equations of motion, the latter depending explicitly  just on the transverse coordinates. Therefore the D(1,4)--D(1,8) brane\footnote{In order to avoid confusions, in Lorentzian signatures we indicate a Dp--brane as a D(1,p)--brane.} system is mapped into a D(5,0)--D(5,4) system  
\be
d\hat{s}_{10}^2=(\tilde{H}_8)^{-\frac12}\left[H_4^{-\frac12} (-dt^2-d\vec{\o}^2)+H_4^{\frac12} (d\vec{x}^2+dz^2)\right]
\ee
We indicate these exotic object as Dirichlet (p,q)--branes as in~\cite{Hull:1998fh} since they are extended objects of dimension p+q, with Neumann boundary conditions in the worldvolume directions and Dirichlet boundary conditions in the transverse directions. 

In order to obtain a solution of ${\rm IIA}_m^*(5,5)$, it is enough to know that ${\rm IIA}_m^*(5,5)$ can also be obtained from ${\rm IIA}_m(5,5)$ by reversing the signature~\cite{Hull:1998ym}. Therefore one immediately obtains  a system of D(0,5) --D(4,5) branes  
\be
d\hat{s}_{10}^2=(\tilde{H}_8)^{-\frac12}\left[H_4^{-\frac12} (dt^2+d\vec{\o}^2)-H_4^{\frac12} (d\vec{x}^2+dz^2)\right]
\ee
where the intersections condition are the same of \eq{interD48} , since on the D4--D8 system  we have to implement both $\sum_{i=1}^{4}\partial_{x_i}\partial_{x_i}\rightarrow -\sum_{i=1}^{4}\partial_{x_i}\partial_{x_i}$ and $\partial_{\tilde{z}}\partial_{\tilde{z}}\rightarrow -\partial_{\tilde{z}}\partial_{\tilde{z}}$. Therefore $\tilde{H}_4$ and $\tilde{H}_8$ can be read from \eq{HD48}. \\
The near horizon geometry is therefore given by
\be
d\hat{s}_{10}^2=\frac94Q_4^\frac12\left(\frac32Q^{\frac12}_8sin\a\right)^{-\frac13}\left[U^2dy^2_{||}-\frac{dU^2}{U^2}-d\O^2_4 \right]
\ee
where $dy^2_{||}=dt^2+d\vec{\o}^2$ and $d\O^2_4=d\a^2+cos^2\a\,d\O^2_3$, which coincides with \eq{m6+4*}.
Note that the five dimensional world--volume of the localised D(0,5)--brane is Euclidean.\\
Therefore if the ${\rm F_4^*}(1,5)$ supergravity can be obtained as the near horizon geometry of a configuration of D(0,5)--D(4,5) branes,  there should be a correspondence between its  $dS_6$ background and the fixed point of an Euclidean Super Yang--Mills theory leaving on the D(0,5)--brane.


\section{Supersymmetry and pseudo--supersymmetry}

Let us briefly comment on the D(0,5) and D(4,5) branes in the context of Domain--Wall/ Cosmology correspondence. 

Stable DW solutions of a system of gravity coupled to scalars, with a potential $V(\phi)$, are generally "fake" supersymmetric~\cite{Freedman:2003ax}. Which means that it is possible to introduce a real "superpotential $W$ such that $V=2[(W^\prime)^2-W^2]$, where $W^\prime=\frac{\d W}{\d\phi}$ and that the DW solution implies the existence of a Killing spinor obeying
\be D_\m\ve+W\g_\m\ve=0\label{KSE}\ee
that is a supersymmetry--like condition.

The mapping between DW and cosmology~\cite{Skenderis:2006fb} implies that a very similar property holds ad well for cosmological solutions with potential $-V(\phi)$. This is implemented introducing a pure imaginary "superpotential $W=i\tilde{W}$ such that $V=-2[(\tilde{W}^\prime)^2-\tilde{W}^2]$ and the cosmological solution implies the existence of a Killing spinor obeying
\be D_\m\ve+i\tilde{W}\g_\m\ve=0\label{PKSE}\ee
This latter condition is called pseudo--supersymmetry. 

The natural question which arises is if it is possible to embed such "fake" (pseudo--) supersymmetric solutions into "true" supergravity solutions.\\
It is clear that it is not possible to embed both solutions in the same supergravity theory. This is because in  supergravity one has to impose reality conditions on the spinors, hence it is immediate to understand that just one between  \eq{KSE} and \eq{PKSE} can be compatible with a given reality condition. \\
As pointed out in~\cite{Bergshoeff:2007cg} one could provide an embedding for each solution if there exist two theories in the same signature in which  the spinors obey different reality conditions, in particular one compatible with \eq{KSE} and the other with \eq{PKSE}.  They also provide an example, that is the D8--brane solution of ${\rm IIA_m}(1,9)$ corresponds to a cosmological solution of ${\rm IIA^*_m}(1,9)$\footnote{See~\cite{Celi:2004st,Zagermann:2004ac} for  examples of the embedding of  "fake" supersymmetric DW into the DW solution in five dimensional supergravities.}.

In the present paper we found another example of this kind of correspondence.\\
Consider in fact the Killing spinor equation for  a ${\rm IIA_m}(1,9)$ D--brane; we can schematically write it as
\be \mathcal{D}_\m\ve+\a_q e^{\frac{q-5}2\phi}\not\!\!{G}^{(q)}\g_\m (\g_{11})^\frac{q}2\ve=0;\quad\quad q=0,\,2,\,4;\quad \a_q\in\mathbb{R} \label{susy}\ee
where $G^0=m$ and $\g_{11}=\g_0\dots\g_{10}$.\\
Equation \eq{susy} has to be consistent with the  Majorana reality condition on $\ve$, which for  ${\rm IIA_m}(1,9)$ is given by
\be \p^\dagger G_I^{-1}=\p^T C_{-}\label{M1}\ee
where  $C_{-}$ is the charge conjugation matrix and $G_I=\g_0$. Consistency can be checked taking into account that for a space--time signature $(t,s)$ we have\footnote{Remember that here we are using the opposite convention for the space--time signature with respect to ~\cite{D'Auria:2002fh,Vaula:2002cn}, therefore \eq{dag1}, \eq{dag2} differ from the ones presented in~\cite{Vaula:2002cn}.} 
\ba
&&(\g_{a_1}\dots\g_{a_n})^T=(-1)^n C_-^{-1} \g_{a_n}\dots\g_{a_1}C_-\label{conj}\\
&&(\g_{a_1}\dots\g_{a_n})^\dagger=(-1)^{ns} G_I^{-1} \g_{a_n}\dots\g_{a_1}G_I\label{dag1}
\ea 
and that for type ${\rm IIA}$ D--branes $q$ is even, therefore $n$ is odd \eq{susy}.\\
Note that \eq{M1} can be generalised to arbitrary space--time signature by defining $G_I$ as the product of all timelike gamma matrices.  Therefore, as discusses in~\cite{Vaula:2002cn}, equation  \eq{susy}  remains valid for ${\rm IIA_m}(5,5)$.

On the other hand, theories like ${\rm IIA_m^*}(1,9)$, ${\rm IIA_m^*}(5,5)$, ${\rm F_4^*}(1,5)$ are characterised by a different reality condition of spinors~\cite{Hull:1998ym,D'Auria:2002fh, Vaula:2002cn}. It is usually referred as "pseudo--Majorana condition"  and can be imposed according to
\be \p^\dagger G_{II}^{-1}=\p^T C_{-}\label{M2}\ee
where $G_{II}$ is the product of all the spacelike gamma matrices. In this case the hermitian conjugate of a product of $n$ gamma matrices becomes 
\be(\g_{a_1}\dots\g_{a_n})^\dagger=(-1)^{n(t-1)} G_{II}^{-1} \g_{a_n}\dots\g_{a_1}G_{II}\label{dag2}\ee

For signatures $(1,9)$ and $(5,5)$, $t-1$ is even, while $s$ is odd. Which means  that  \eq{dag1} and \eq{dag2} differ for a sign, therefore  for  ${\rm IIA_m^*}(1,9)$ and  ${\rm IIA_m^*}(5,5)$, the Killing spinor equation for a D--brane is given by 
\be \mathcal{D}_\m\ve\pm i\,\a_q e^{\frac{q-5}2\phi}\not\!\!{G}^{(q)}\g_\m (\g_{11})^\frac{q}2\ve=0;\quad\quad q=0,\,2,\,4;\quad \a_q\in\mathbb{R} \label{susy*}\ee
where the $\pm$ sign depends on how the RR fields are redefined \eq{map10} and has no consequences on the reality condition. This is the generalisation of the DW/Cosmology map $iW=\tilde{W}$, which we retrieve for the case $q=0$, corresponding to the example in~\cite{Bergshoeff:2007cg}. 

In conclusion, the observation in~\cite{Bergshoeff:2007cg} that pseudo--supersymmetry in ${\rm IIA_m}(1,9)$ cosmologies corresponds to supersymmetry in ${\rm IIA_m^*}(1,9)$ is actually more general. In particular, it is possible to check if given a fake supersymmetric  solution, supported by a field strength $G^{(q)}$, which can be embedded into a real supergravity theory, the corresponding fake pseudo--supersymmetric  solution can be embedded into a variant theory. It is in fact sufficient to check using \eq{conj},   \eq{dag1} and \eq{dag2}  if the fake pseudo Killing spinor equation can be obtained from the supersymmetry variation of a variant theory, in the same spirit of~\cite{Vaula:2002cn}. Note that it is not excluded that the corresponding variant theory has a different space--time signature. This seems to be the case e.g. for M--branes in eleven dimensions and for NS--branes in ten dimensions, since a change of reality in the three--form $C_{(3)}\rightarrow iC_{(3)}$ and in NSNS three--form $H_{(3)}\rightarrow iH_{(3)}$ respectively, are always associated with a change in the space--time signature~\cite{Hull:1998ym,Vaula:2002cn}

\begin{center}
\bf{Acknowledgments}
\end{center}     
The author would like to thank Patrick Meessen,Tom\'as Ort\'\i n and Jos\'e Barb\'on for their valuable help and criticism.
 This work has been supported in part by the Spanish Ministry of Science and
Education grant FPA2006-00783, the Comunidad de Madrid grant HEPHACOS
P-ESP-00346 and by the EU Research Training Network \textit{Constituents,
  Fundamental Forces and Symmetries of the Universe} MRTN-CT-2004-005104.

\appendix

\section{}
For completeness we write the compactification ansatz of ${\rm IIA_m^*}(5,5)$. This can be obtained applying the maps \eq{map10}, \eq{map6} to the ansatz in~\cite{Cvetic:1999un}.
\ba
d\hat s_{10}^2\!\! &=&\!\! (\sin\zeta)^{\frac1{12}} X^{\frac18}\Big[
\Delta^{\frac38}\, ds_6^2 - 2g^{-2} \Delta^{\frac38} X^2\, d\zeta^2
-\frac12g^{-2}\Delta^{-\frac58} X^{-1} \cos^2\zeta
\sum_{i=1}^3(\sigma^i - g\, A_{(1)}^i)^2\Big]\,,\nn\\
\hat F_{(4)} &=& -\frac{\sqrt2}{6}\, g^{-3}\, s^{1/3}\, c^3\, \Delta^{-2}\,
U\, d\zeta\wedge\e_{(3)} -\sqrt2 g^{-3}\, s^{4/3}\, c^4\, \Delta^{-2}\,
X^{-3}\, dX\wedge \e_{(3)} \nn\\
&&-\sqrt2 g^{-1}\, 
s^{1/3}\, c\, X^4\, {*H_{(3)}}\wedge d\zeta
+\frac1{\sqrt2} s^{4/3}\, X^{-2}\, {*F_{(2)}} \nn\\
&& +\frac1{\sqrt2} g^{-2}\,
s^{1/3}\, c\, F_{(2)}^i \, h^i\wedge d\zeta -\frac1{4\sqrt2} g^{-2}\,
s^{4/3}\, c^2\, \Delta^{-1}\, X^{-3}\,  F_{(2)}^i \wedge
h^j\wedge h^k\, \e_{ijk}\,,\label{fans}\\
\hat H_{(3)} &=& s^{2/3}\, H_{(3)} + g^{-1}\, s^{-1/3}\, c\, F_{(2)}\wedge d\zeta
\,,\nn\\
\hat F_{(2)} &=& \frac1{\sqrt2}\, s^{2/3}\, F_{(2)}\,,\qquad
e^{\hat\phi} = s^{-5/6}\, \Delta^{1/4}\, X^{-5/4}\,,\nn
\ea
where $X$ is related to the four dimensional dilaton $\phi$ by
$X=e^{-\frac1{2\sqrt2}\phi}$, and
\ba
\Delta &\equiv & X\cos^2\zeta +X^{-3} \sin^2 \zeta\,,\nn\\
U &\equiv& X^{-6}\, s^2 - 3 X^2\, c^2 + 4 X^{-2}\, c^2 - 6 X^{-2}\,.
\ea
Furthermore, the functions $h$, $f_1$, $f_2$ introduced for brevity in \eq{m10} are given by
\be h(\z,\phi)=  (\sin\zeta)^{\frac1{12}}\, X^{\frac18}\Delta^{\frac38};\quad f_1(\z,\phi)=X^2;\quad  f_2(\z,\phi)=\frac14 \D^{-1}X^{-1} \ee
while $\e_{(3)}\equiv h^1\wedge h^2\wedge h^3$, and $s=\sin\zeta$ and $c=\cos\zeta$.  The gauge
coupling constant $g$ is related to the mass parameter $m$ by $m= \frac{\sqrt2}{3}\, g$.

Note, that the compactification ansatz  differs from the one in~\cite{Cvetic:1999un} just for a few signs, which take into account the corresponding modifications of the equations of motion. 

\appendixend


\begin{thebibliography}{99}

\bibitem{Romans:1985tw}
  L.~J.~Romans,
  Nucl.\ Phys.\  B {\bf 269} (1986) 691.

\bibitem{VanProeyen:1999ni}
  A.~Van Proeyen,
  arXiv:hep-th/9910030.

\bibitem{Ferrara:2001dn}
  S.~Ferrara and M.~A.~Lledo,
  Rev.\ Math.\ Phys.\  {\bf 14} (2002) 519
  [arXiv:hep-th/0112177].

\bibitem{D'Auria:2000ad}
  R.~D'Auria, S.~Ferrara and S.~Vaula,
  JHEP {\bf 0010} (2000) 013
  [arXiv:hep-th/0006107].

\bibitem{Andrianopoli:2001rs}
  L.~Andrianopoli, R.~D'Auria and S.~Vaula,
  JHEP {\bf 0105} (2001) 065
  [arXiv:hep-th/0104155].



\bibitem{D'Auria:2002fh}
  R.~D'Auria and S.~Vaula,
  JHEP {\bf 0209} (2002) 057
  [arXiv:hep-th/0203074].


\bibitem{Hull:1998vg}
  C.~M.~Hull,
  JHEP {\bf 9807} (1998) 021
  [arXiv:hep-th/9806146].


\bibitem{Cvetic:1999un}
  M.~Cvetic, H.~Lu and C.~N.~Pope,
  ``Gauged six-dimensional supergravity from massive type IIA,''
  Phys.\ Rev.\ Lett.\  {\bf 83} (1999) 5226
  [arXiv:hep-th/9906221].


\bibitem{Romans:1985tz}
  L.~J.~Romans,
  Phys.\ Lett.\  B {\bf 169} (1986) 374.


\bibitem{Hull:1998ym}
  C.~M.~Hull,
  JHEP {\bf 9811} (1998) 017
  [arXiv:hep-th/9807127].



\bibitem{Ferrara:1998gv}
  S.~Ferrara, A.~Kehagias, H.~Partouche and A.~Zaffaroni,
  Phys.\ Lett.\  B {\bf 431} (1998) 57
  [arXiv:hep-th/9804006].



\bibitem{Brandhuber:1999np}
  A.~Brandhuber and Y.~Oz,
  Phys.\ Lett.\  B {\bf 460} (1999) 307
  [arXiv:hep-th/9905148].


\bibitem{Skenderis:2006fb}
  K.~Skenderis and P.~K.~Townsend,
  arXiv:hep-th/0610253.


\bibitem{Bergshoeff:2007cg}
  E.~A.~Bergshoeff, J.~Hartong, A.~Ploegh, J.~Rosseel and D.~Van den Bleeken,
  arXiv:0704.3559 [hep-th].



\bibitem{Skenderis:2007sm}
  K.~Skenderis, P.~K.~Townsend and A.~Van Proeyen,
  arXiv:0704.3918 [hep-th].



\bibitem{Liu:2003qa}
  J.~T.~Liu, W.~A.~Sabra and W.~Y.~Wen,
  JHEP {\bf 0401} (2004) 007
  [arXiv:hep-th/0304253].


\bibitem{Vaula:2002cn}
  S.~Vaula,
  ``On the construction of variant supergravities in D = 11, D = 10,''
  JHEP {\bf 0211} (2002) 024
  [arXiv:hep-th/0207080].


 
\bibitem{Youm:1999zs}
  D.~Youm,
  arXiv:hep-th/9902208.




\bibitem{Hull:1998fh}
  C.~M.~Hull and R.~R.~Khuri,
  Nucl.\ Phys.\  B {\bf 536} (1998) 219
  [arXiv:hep-th/9808069].





\bibitem{Freedman:2003ax}
  D.~Z.~Freedman, C.~Nunez, M.~Schnabl and K.~Skenderis,
  Phys.\ Rev.\  D {\bf 69} (2004) 104027
  [arXiv:hep-th/0312055].




\bibitem{Celi:2004st}
  A.~Celi, A.~Ceresole, G.~Dall'Agata, A.~Van Proeyen and M.~Zagermann,
  Phys.\ Rev.\  D {\bf 71} (2005) 045009
  [arXiv:hep-th/0410126].


\bibitem{Zagermann:2004ac}
  M.~Zagermann,
  Phys.\ Rev.\  D {\bf 71} (2005) 125007
  [arXiv:hep-th/0412081].




\end{thebibliography}
\end{document}